\documentclass[letterpaper,12pt]{article}   

\usepackage{osajnl2} 
\usepackage{graphics, epsfig}

\begin{document}

\title{Light propagation through a coiled optical fiber and Pancharatnam phase
}


\author{\bf Rajendra~Bhandari}
\address{  Raman Research Institute, \\ Bangalore 560 080, India. \\ email: bhandari@rri.res.in}

\begin{abstract}The nature of changes in the interference pattern caused by the presence of polarization-changing elements   in one or both beams of an interferometer, in particular those caused by an effective optical activity due to passage of a polarized beam through a coiled optical fiber are clarified. It is pointed out that for an incident state that is not circularly polarized so that the two interfering beams go to different polarization states, there is an observable nonzero Pancharatnam phase shift between them which depends on the incident polarization state and on the solid angle subtended by the track of the $\vec{k}$-vector at the centre of the sphere of $\vec{k}$-vectors. The behaviour of this phase shift is singular when the two interfering states are nearly orthogonal. It is shown that  for zero path difference between the two beams,  the amplitude of intensity modulation as a function of  optical activity is independent of the incident polarization state. 
\end{abstract}

\ocis{260.3160, 260.5430, 060.2310, 350.1370}

\maketitle 

\section{Introduction}

The nature of changes in an interference pattern resulting from a change in the 
polarization of one or both beams in an interferometer have received considerable 
attention in recent years following the work of Berry \cite{berry1} in 
which he discovered an interesting phase-shift effect on the wavefunction of a 
quantum mechanical system evolving under the action of a cyclic, adiabatically varying 
hamiltonian. Having their origin in the work of  Pancharatnam \cite{panch1}, 
topological phase shifts arising from polarization transformations in a light 
wave with a fixed $\vec{k}$-vector were measured in interference experiments and shown 
to have many counter-intuitive properties \cite{unbounded,jumps,4pism,iwbs,rbdirac,rbphysica,rbreview}.
Another interesting topological phase shift, arising from a cyclic change in 
the direction of propagation of a light beam along a three-dimensional curve in space, for 
example that caused by the passage of a monochromatic beam through a monomode optical fibre, 
was predicted by Chiao and Wu \cite{chiaowu}. The basic effect here is the parallel 
transport of the polarization of the wave as it traverses a three-dimensional 
curve in space such that the tip of its $\vec{k}$-vector goes around a closed curve on 
the sphere of $\vec{k}$-vectors. In such a transport, a circularly polarized wave
propagates without a change in its polarization state but acquires a geometric phase shift 
$\pm\gamma$, where $\gamma$ is equal to the solid 
angle subtended by the track of the $\vec{k}$-vector at the centre of 
the sphere of $\vec{k}$-vectors; the sign of the phase shift being different for the two different 
circular polarizations. This has been known in literature as the "spin redirection phase". 
The difference between the phases acquired by the left and the right
circularly polarized states in such a transport was experimentally demonstrated by
Frins and Dultz  \cite{frinsdultz} to be of magnitude $2\gamma$.  

An arbitrarily polarized state propagating 
through the fiber, however, experiences a rotation of its polarization ellipse 
about the $\vec{k}$-vector through an angle equal to $\gamma$.   This corresponds to a 
rotation of the representative point   on the 
Poincar\'{e} Sphere about the polar axis of the 
sphere through an angle $2\gamma$; the poles representing the circularly polarized states.  
Such a rotation, for linearly polarized states, was experimentally demonstrated 
by Tomita and Chiao \cite{tomchiao}. This effect, originating in the degeneracy of the 
two circularly polarized states, was interpreted by Anandan \cite{anandan} as an 
example of the Wilczek-Zee phase \cite{wilczek}. This can be looked upon as an 
effective optical activity.
A question may be asked : in addition to such a rotation, which represents a change in 
the {\it polarization state} of the beam, does the beam experience a phase shift ? The answer 
is yes and this is the main subject of this paper. The omission  of this phase shift 
may lead to incorrect conclusions about the results of interference experiments. An example is 
a recent paper by Senthilkumaran \cite{senthil3} in which a related experimental
situation namely a "tunable fibre optic mirror" is analyzed and  the absence of such a 
phase shift is assumed. The conclusions arrived at contradict the results of earlier 
experimental work on the tunable fiber optic mirror \cite{senthil1,senthil2}.
Since interference situations where the polarization states of one or both beams 
change are commonly encountered and since these are not  treated correctly in existing 
books on interferometry (for example see ref.\cite{hariharan}), we attempt to present, in the 
following sections, a correct analysis of the situation.

\section{Interference of polarized light}

In this section we present an analysis of  interference of 
polarized light in the presence of polarization-changing elements in the path of 
one or both interfering beams in the context of the simplest interference situation 
namely Young's two-slit interference experiment (fig.1).

\begin{figure*}
\centerline{
\epsfxsize=0.7\textwidth
\epsfbox{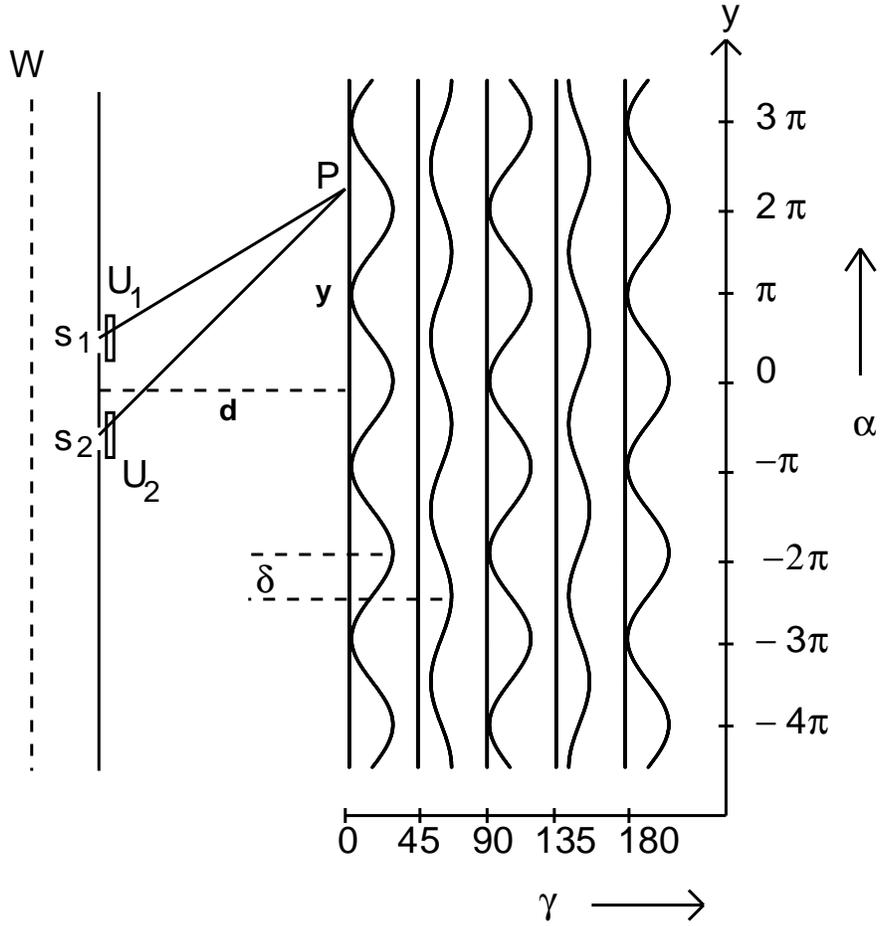}
}
\caption{This figure shows the interference pattern on a screen formed by interference of two parts of a wavefront W through  slits $S_1$ and $S_2$ which have undergone polarization transformations $J_1$ and $J_2$ corresponding to an 
effective optical activity that causes a rotation about the beam axis through angles $\gamma$ and $-\gamma$ respectively. Intensity variation on the screen as a function of the distance $y$ along the screen (approximately proportional to the optical path difference $\alpha$ between the two beams) is shown for 5 different values of $\gamma$, i.e $\gamma=0^\circ$, $\gamma=45^\circ$, $\gamma=90^\circ$, $\gamma=135^\circ$ and $\gamma=180^\circ$. As $\gamma$ changes, the visibility of the fringes changes and the fringes shift along the $y$-axis by an amount $\delta$ which is the Pancharatnam phase difference between the beams. Note, when $\alpha=n\pi$, the amplitude of the intensity modulation as a function of $\gamma$ is 1, whereas for an arbitrary  value of $\alpha$, the modulation can be less than 1. The fringes shown in the figure correspond to a polarization 
state with $\theta=60^\circ$.}
\label{fig.1}
\end{figure*}

Since our focus in this paper is on optical activity, we shall use  circularly polarized 
states as the basis states unless stated otherwise. The variable $n$ is an integer throughout the paper.

A unit intensity beam in polarization state $\eta$ can be represented  by a two-component complex column vector

\begin{equation}
\eta = \left( 
  \begin{array}{l}
{\rm cos}(\theta/2) \\
{\rm sin}(\theta/2){\rm exp}(i\phi) \\
\end{array}
\right)
\end{equation}

\noindent where $\theta$ and $\phi$ are the polar and azimuthal coordinates respectively of the point 
representing the polarization state on the Poincar\'{e} Sphere.
To keep the discussion general at this point let the slits $s_1$ and $s_2$ have different widths so that the 
beams exiting $s_1$ and $s_2$ have different intensities.
Beams with intensities $I_1$ and $I_2$ in polarization state $\eta$ can be represented  by the 
complex column vectors

\begin{equation}
{\eta}_1 =\sqrt{I_1} \eta~~{\rm and}~~\\
{\eta}_2 =\sqrt{I_2} \eta
\end{equation}

\noindent Let the beams after passing through the slits $s_1$ and $s_2$ pass through a box placed in front of 
each of the slits where  unitary transformations $J_1$ and $J_2$ are performed on the states ${\eta}_1$
and ${\eta}_2$ respectively following which the beams travel to a point P on the screen where 
the intensity resulting from their interference is considered (fig.1). The transformations  $J_1$ and $J_2$
can each be broken into two factors, the first representing an  isotropic (polarization-independent) part 
that multiples the 
state vectors by phase factors ${\rm exp}(i\beta_1)$ and ${\rm exp}(i\beta_2)$  and the second, 
an SU(2) part that multiplies the state vectors by  2 x 2 complex unitary matrices $U_1$ and $U_2$ with 
determinant +1. 
So that,

\begin{equation}
J_1 = {\rm exp}(i\beta_1) U_1 ~~{\rm and}~~  J_2 = {\rm exp}(i\beta_2) U_2
\end{equation}

\noindent The states $\psi_1$ and $\psi_2$ at P are  given by, 

\begin{eqnarray}
 & &\psi_1  =  J_1~ {\rm exp}(i\alpha_1)~\eta_1 ~~ =\sqrt {I_1}~{\rm exp}i(\beta_1+\alpha_1)~\tilde{\psi_1} ~~{\rm and}~~ \nonumber \\
 & &\psi_2  =  J_2~ {\rm exp}(i\alpha_2)~\eta_2 ~~ = \sqrt {I_2}~{\rm exp}i(\beta_2+\alpha_2)~\tilde{\psi_2} 
\end{eqnarray}

\noindent where

\begin{eqnarray}
\tilde{\psi_1} = U_1~\eta,~ \tilde{\psi_2} = U_2~\eta ~~{\rm and}~~ 
{{\tilde{\psi_2}}^{\dagger}}\tilde{\psi_2}={{\tilde{\psi_1}}^{\dagger}}\tilde{\psi_1}=1 
\end{eqnarray}

\noindent The phase factors ${\rm exp}(i\alpha_1)$ and ${\rm exp}(i\alpha_2)$ are due to propagation of the 
two waves from 
the slits to the point P.

The intensity $I$ at P is given by,

\begin{eqnarray}
 & &I = {\psi_1}^{\dagger}{\psi_1}~+~{\psi_2}^{\dagger}{\psi_2}~+2Re[{\psi_2}^\dagger{\psi_1}]\nonumber\\
 & &=I_1+I_2+2(I_1I_2)^{\frac{1}{2}}Re[{\tilde{\psi_2}}^\dagger\tilde{\psi_1}{\rm exp}i(\beta+\alpha)]\nonumber\\
 & &=I_1+I_2+2(I_1I_2)^{\frac{1}{2}}Re[\xi {\rm exp}i(\delta+\beta+\alpha)]\label{eq:intensity}\\
& &{\rm where} ~~ \beta=\beta_1-\beta_2 ,~~ \alpha=\alpha_1-\alpha_2 ~~{\rm and}~~
{\tilde{\psi_2}}^\dagger\tilde{\psi_1}=\xi {\rm exp}(i\delta)  
\end{eqnarray}

\noindent The phase difference $\alpha$ due to the waves from slits 1 and 2 having travelled different path lengths 
to the point P is given, for s/d$<<$1 and y/d$<<1$ by,

\begin{equation}
\alpha=(2\pi {\rm s}/\lambda)({\rm y/d})
\end{equation}

\noindent where $\lambda$ is the wavelength of light, s is the distance between the slits, d is the 
distance between the planes containing the slits and the screen and y is the distance along the screen 
of the point P  
from the point of zero path difference between the waves.
We define the quantity $\delta$=arg$({\tilde{\psi_2}}^\dagger\tilde{\psi_1})$
=arg$({\eta}^\dagger {U_2}^\dagger U_1 \eta)$ to be the Pancharatnam phase 
shift between the two beams due to the SU(2)transformations $U_1$ and $U_2$. We have deliberately 
kept the isotropic phase shifts $\beta_1$ and $\beta_2$ out of the definition of the Pancharatnam 
phase so that $\delta$ represents the effect of polarization transformations alone.
The intensity given by eqn.(\ref{eq:intensity}) has maxima when $(\delta+\beta+\alpha)=2n\pi$ 
and minima when $(\delta+\beta+\alpha)=(2n+1)\pi$. These are given by,

\begin{eqnarray}
 & &I^{max}= I_1+I_2+2(I_1I_2)^{\frac{1}{2}}\xi  ~~ and \\
 & &I^{min}= I_1+I_2-2(I_1I_2)^{\frac{1}{2}}\xi
\end{eqnarray}

\noindent so that the visibility $V$ is given by,

\begin{equation}
V=(I_{max}-I_{min})/(I_{max}+I_{min})=[2(I_1I_2)^{\frac{1}{2}}/(I_1+I_2)]\xi
\end{equation}

\noindent when $I_1=I_2$, $V=\xi$. The quantity $\xi=\mid {\tilde{\psi_2}}^\dagger\tilde{\psi_1}\mid$ 
is then a measure of the visibility of the interference fringes. The  transformations $J_1$ and $J_2$
thus have two effects on the interference pattern: (i) the contrast of the pattern changes 
from 1 to $\xi$ and (ii) the fringes shift along the $y$-axis (or say the $\alpha$-axis) 
by an amount $-(\beta+\delta)$.

Let us next consider the case when the SU(2) transformations $U_1$ and $U_2$ correspond to 
optical activity so that the incident polarization states $\eta_1$ and $\eta_2$ are 
rotated about the beam axis through angles $\gamma$ and $-\gamma$ respectively without any 
change in the ellipticity of the polarization ellipse. Let us also assume that $\beta_1=\beta_2=0$.
Examples of such transformations are (i) passage of a polarized beam through a coiled optical fibre 
with integral number of turns and (ii) passage of a polarized beam with a fixed $\vec k$ vector 
through a pair of halfwave plates whose principal axes make an angle $\gamma/2$ with each other.
We then have,

\begin{eqnarray}
 & & J_1=U_1=\left( 
  \begin{array}{lr}
{\rm exp}(-i\gamma) & 0 \\
0 &{\rm exp}(i\gamma) \\
\end{array}
\right) ~{\rm and}~ \nonumber \\
 & & J_2=U_2=\left( 
  \begin{array}{lr}
{\rm exp}(i\gamma) & 0 \\
0 &{\rm exp}(-i\gamma) \\
\end{array}
\right) 
\end{eqnarray}

\noindent so that,

\begin{equation}
{\tilde{\psi_2}}^\dagger\tilde{\psi_1}={\eta}^\dagger {U_2}^\dagger U_1 \eta=\xi {\rm exp}(i\delta)\\
={\rm cos}^2 (\theta/2) {\rm exp}(2i\gamma)+{\rm sin}^2 (\theta/2) {\rm exp}(-2i\gamma)\label{eq:visibility}
\end{equation}

\noindent This gives,

\begin{equation}
\xi {\rm cos}\delta = {\rm cos}(2\gamma) ~~{\rm and}~~ \xi {\rm sin}\delta = {\rm cos}\theta sin(2\gamma) \label{eq:phase1}
\end{equation}

\noindent or,

\begin{equation}
\xi=[{\rm cos}^2 (2\gamma)+{\rm cos}^2 (\theta){\rm sin}^2 (2\gamma)]^{\frac{1}{2}} ~~{\rm and}~~ {\rm tan}\delta={\rm cos}\theta {\rm tan}(2\gamma) \label{eq:phase2}
\end{equation}

The visibility $\xi$ and the Pancharatnam phase shift $\delta$ can be determined from the pair 
of equations (\ref{eq:phase1}) or (\ref{eq:phase2}). Figure 1 shows the intensity variation on 
the screen along the y-axis for 5 different values of $\gamma$ for an incident state corresponding 
to $\theta=60^\circ$ and for $I_1=I_2=1/2$. As $\gamma$ changes, there are two changes in the interference pattern. 
The amplitude of the intensity variation, i.e. the visibility $\xi$ given by eqn.(\ref{eq:phase2}) 
changes and the intensity curve shifts along the y-axis by an amount $\delta$, also given 
by eqn.(\ref{eq:phase2}) or eqn.(\ref{eq:phase1}). In fig.1, $\xi=1$ for $\gamma=0^\circ, 90^\circ, 180^\circ$ 
and  $\xi=1/2$ for $\gamma=45^\circ, 135^\circ$.
The integrated phase shift $\int {d\delta}$ 
determined from these equations for a typical set of values of the polar angle $\theta$ of the 
incident polarization state on the Poincar\'{e} Sphere are shown in fig.2.

\begin{figure*}
\centerline{
\epsfxsize=0.5\textwidth
\epsfbox{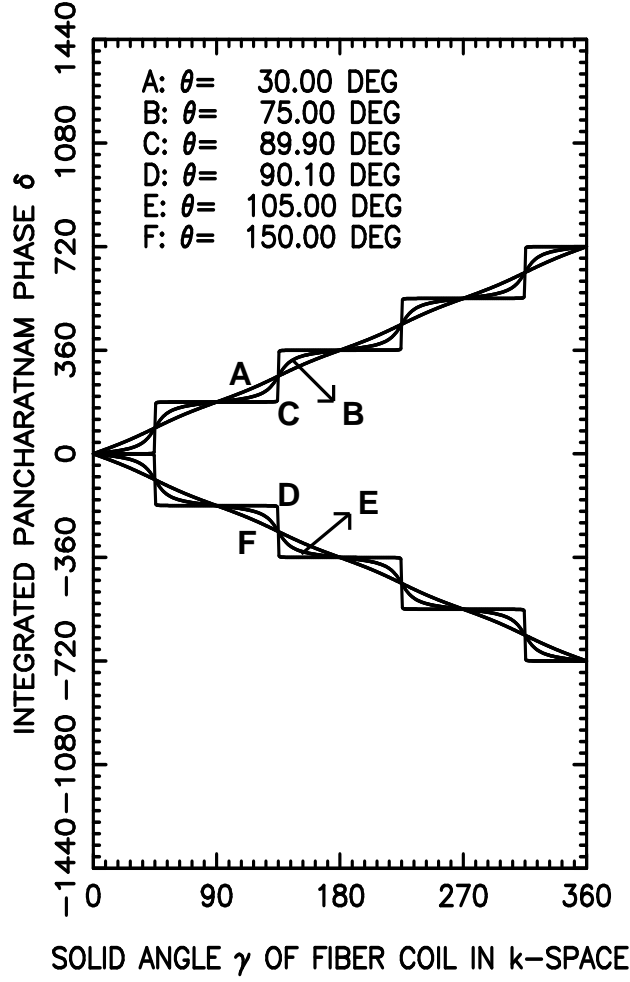}
}
\caption{This figure shows the Pancharatnam phase shift $\delta$ in degrees as a function of the optical activity parameter $\gamma$ which in the case of propagation through a fiber loop is equal to the solid angle subtended by the track of the $\vec{k}$-vector at the centre of the sphere of $\vec{k}$-vectors. The 6 curves A, B, C, D, E and F correspond to incident polarization states with polar angle $\theta=30^\circ, 75^\circ, 89.9^\circ, 90.1^\circ, 105^\circ$ and $150^\circ$. The curves for $\theta=0^\circ$ 
and $\theta=180^\circ$ are straight lines nearly coincident with the curves A and F  respectively and are not shown separately.
For $\gamma=n\pi/2$ sterradians,   the polarization of the two beams undergoes rotation through $\pi$ and $-\pi$ on the Poincar\'{e} sphere and the total phase shift has magnitude $\pi$ irrespective of $\theta$. Also note the singular behaviour for $\theta=90^\circ$when $\gamma$ has the values $(2n+1)\pi/4$ sterradians.}
\label{fig.2}
\end{figure*}
 
The curves in fig.2 
represent the phase shift that would be measured by an interferometer that can keep track of 
the phase shift continuously as the parameter $\gamma$ 
representing the optical activity is changed. As seen from equations (\ref{eq:phase1}) 
and (\ref{eq:phase2}) and from fig.2, at the values $(\theta=90^\circ,\gamma=(2n+1)(\pi /4))$, 
the visibility $\xi$ of the interference pattern becomes zero and the phase shift becomes singular.
The two interfering states are orthogonal at these points. 
The phase varies sharply near these points with an abrupt change in sign and a closed circuit in the parameter space $(\gamma,\theta)$
around one of these points results in a total integrated phase shift equal to $\pm 2n\pi$.
Such a behaviour of the phase shifts caused by polarization transformations was first predicted 
using a gedanken polarization experiment in ref.\cite{jumps} and has been observed in 
interference experiments using quarterwave and halfwave retarders as the 
SU(2) elements \cite{4pism,iwbs,rbdirac}. 
In the context where $\gamma$ originates 
in transport through a coiled 
optical fibre, the phase shift shown in fig.2 corresponding to $\theta=0$ has been seen 
in experiments reported by Frins and Dultz \cite{frinsdultz}. A simple extension of this 
experiment that allows the incident polarization state to be varied 
would enable the full behaviour of the phase shift to be observed except very near the 
singularities where the interference contrast becomes too low.
Let us note that for $\gamma=90^\circ$, the polarization states of one of the two beams rotates
through $\pi$ on the Poincar\'{e} sphere and that of the other beam rotates through $-\pi$. 
For this relative rotation of $2\pi$, the total phase shift is of magnitude $\pi$ irrespective of 
the value of $\theta$. This is analogous to the sign change of spin-1/2 wavefunctions under 
$2\pi$ rotations in real space. It may also be pointed out that the phase shift defined as 
above is the total phase shift due to the SU(2) transformation and not just the geometric 
part of the phase. This is elaborated further in section 5.

\section{The tunable fiber optic mirror}
In refs.\cite{senthil1,senthil2}, Senthilkumaran et al. reported a fiber optic device 
in which they use a coiled fiber in a Sagnac interferometer configuration in 
combination with a halfwave retarder to produce equal and opposite phase shifts $\gamma$
and $-\gamma$
on the propagating and counter-propagating circularly polarized beams respectively which result in a
complimentary modulation of intensity at the two output ports of the interferometer 
as a function of the phase shift introduced; the latter being equal to the solid angle 
subtended by the track of the tangent vector to the fiber coil at the centre of the 
sphere of directions in space. It was observed in these experiments that when an 
arbitrary linear superposition of the two circularly polarized states is incident on 
the interferometer, the  
modulation index of intensity at the output port is independent of the incident polarization state. In this 
section we try to understand this result in the light of the discussion in the 
previous section.

Let a halfwave retarder oriented with its fast axis making an angle $\tau$ with the 
x-axis be represented by $H(\tau)$ and an optical rotator that rotates any  polarization 
through an angle $\gamma$ about the beam axis be represented by $R(\gamma)$.
In the Sagnac configuration used in refs.\cite{senthil1,senthil2}, the propagating beam 
undergoes an SU(2) transformation $U_1$ corresponding to a product of $H(0)$ and $R(\gamma)$
i.e. 

\begin{eqnarray}
 & &U_1=H(0)R(\gamma)=-i\left( 
  \begin{array}{lr}
0 & 1 \\
1 & 0 \\
\end{array}
\right) 
\left( 
  \begin{array}{lr}
{\rm exp}(-i\gamma) & 0 \\
0 &{\rm exp}(i\gamma) \\
\end{array}
\right) \nonumber \\
& & =-i\left( 
  \begin{array}{lr}
0 & {\rm exp}(i\gamma) \\
{\rm exp}(-i\gamma) & 0\\
\end{array}
\right) 
\end{eqnarray}

\noindent The counterpropagating beam on the other hand sees a transformation $U_2$ 
which is a product of the same two transformations in the 
reverse order, i.e. 

\begin{eqnarray}
U_2=R(\gamma)H(0)=
-i\left( 
  \begin{array}{lr}
0 & {\rm exp}(-i\gamma) \\
{\rm exp}(i\gamma) & 0\\
\end{array}
\right) 
\end{eqnarray}

\noindent We have assumed as before that $\beta_1=\beta_2=0$. Let us now assume that
a unit intensity is incident on the beam splitter of the interferometer and that 
the beam splitter divides the intensity between the transmitted and  reflected 
waves in the ratio $1:1$ so that $\eta_1=\eta_2=\eta$. Since the beam splitter 
acts twice before the beams exit the output ports, $I_1=I_2=1/4$ and the intensity $I$ 
at one of the output ports
is  given , following eqn.(\ref{eq:intensity}), by

\begin{eqnarray}
I = (1/2)[1+Re({\tilde{\psi_2}}^\dagger\tilde{\psi_1}{\rm exp}i\alpha)]\label{eq:intensity2}
\end{eqnarray}

\noindent where

\begin{equation}
{\tilde{\psi_2}}^\dagger\tilde{\psi_1}={\eta}^\dagger {U_2}^\dagger U_1 \eta=\xi {\rm exp}(i\delta)\\
={\rm cos}^2 (\theta/2) {\rm exp}(-2i\gamma)+{\rm sin}^2 (\theta/2) {\rm exp}(2i\gamma)\label{eq:visibility2}
\end{equation}

\noindent Notice the similarity of eqn.(\ref{eq:visibility2}) with eqn.(\ref{eq:visibility}). Except for the 
sign of $\gamma$, the two expressions are the same. The interference in this case is also 
between two states with the same ellipticity, but rotated with respect to each other about 
the beam axis through an angle $2\gamma$. However in this case due to the presence of H(0) 
there is a change in the 
helicity of both the states with respect to the initial state which also accounts for the 
change in the sign of $\gamma$. Therefore the Pancharatnam phase in this case also  has 
the same general behaviour as shown in fig.2, with the appropriate sign changes.

Let us substitute  eqn.(\ref{eq:visibility2}) in eqn.(\ref{eq:intensity2}).
This gives, 

\begin{eqnarray}
 & &I = (1/2)[1+\xi {\rm cos}(\delta+\alpha)]\nonumber\\
 & &=(1/2)[1+{\rm cos}^2(\theta/2) {\rm cos}(\alpha-2\gamma)+{\rm sin}^2(\theta/2) {\rm cos}(\alpha+2\gamma)]\nonumber \\
 & & =(1/2)[1+{\rm cos}\alpha {\rm cos}(2\gamma)+{\rm sin}\alpha {\rm sin}(2\gamma) {\rm cos}\theta] \label{eq:intensity3}
\end{eqnarray}

\noindent Rewriting the right hand side in eqn.(\ref{eq:intensity3}),

\begin{eqnarray}
 & & I = (1/2)[1+M {\rm cos}(2\gamma-\chi)],~~{\rm where}\nonumber\\
 & & M{\rm cos}\chi={\rm cos}\alpha,\nonumber\\
 & & M{\rm sin}\chi={\rm cos}\theta {\rm sin}\alpha ~~{\rm and} \nonumber\\
 & & M=[{\rm cos}^2 \alpha+{\rm cos}^2 \theta {\rm sin}^2 \alpha]^{\frac{1}{2}} \label{eq:modulation}
\end{eqnarray}

\noindent Eqn.(\ref{eq:modulation}) is the main result of this section and shows that at points in the interference pattern 
where $\alpha=0$ or $n\pi$, $M=1$.
This result is independent of $\theta$, i.e. of the incident polarization state.
This is indeed what was observed in refs.\cite{senthil1,senthil2}, where the path difference $\alpha$ between the two intefering beams is zero.
The conclusion  in ref.\cite{senthil3} that $M$ depends on $\theta$ results from the assumption therein
that $\delta=0$, which is incorrect.
When $\theta=0$ or $\pi$, i.e. when the incident state is circularly polarized, eqn.(\ref{eq:modulation})
gives,

\begin{equation}
I=(1/2)[1 + {\rm cos}(2\gamma \pm \alpha)], \label{eq:modulation2}
\end{equation}

\noindent so that $M$ in this case is 1 irrespective of the 
value of $\alpha$. In ref.\cite{senthil3} it was 
concluded that $M=0$ for this case.  For an arbitrary 
$\theta$, however, $M$ is a function of $\alpha$,  being $1$ for $\alpha=0$ as in refs.\cite{senthil1,senthil2}. 

A simple equivalent of the tunable fiber optic mirror in the context of conventional, 
non-fiber interferometry is shown in fig.3. BS is a 50:50 beam splitter that reflects incident 
circularly polarized light with a change of helicity  and $M_1, M_2, M_3$ are mirrors that 
do the same. Modulation of the intensity at either 
of the two output ports in this case can be achieved by rotation of the halfwave plate H about the beam axis;
rotation of H through an angle $\gamma/2$ being equivalent to the solid angle $\gamma$ of the fiber loop.
This equivalence can easily be seen if we note that (i) optical activity is equivalent to passage through a 
pair of halfwave plates with their principal axes making a certain angle with each other and 
(ii) any sequence of three halfwave plates is equivalent to a single halfwave plate oriented at 
an appropriate angle \cite{rbphysica}.

\begin{figure*}
\centerline{
\epsfxsize=0.6\textwidth
\epsfbox{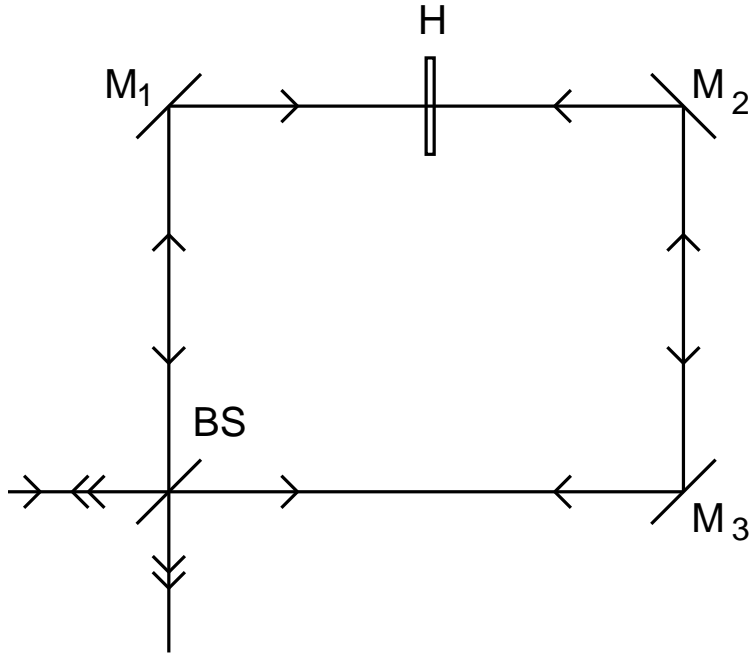}
}
\caption{This figure shows a conventional Sagnac interferometer configuration that would act as a tunable mirror equivalent to the tunable fiber optic mirror. 
A rotation of the halfwave plate $H$  about the beam axis results in a complimentary modulation of  intensity at the two output ports. Rotation through an angle $\gamma/2$ is equivalent to passage through a fiber loop with solid angle $\gamma$ in $\vec k$-space.}
\label{fig.3}
\end{figure*}

\section{A pair of halfwave plates}

A pair of aligned halfwave plates, i.e. a fullwave plate is an interesting device in 
that {\it any} polarization state passing through it acquires a topological phase shift of 
magnitude $\pi$ in addition to any isotropic phase shift due to an extra path length. 
In the basis of circularly polarized states, the Jones matrix (the SU(2) part) of a 
halfwave plate with its fast axis making an 
angle $\tau/2$ with the x-axis is given by,

\begin{eqnarray}
H(\tau/2)=-i
\left( 
  \begin{array}{lr}
0 & {\rm exp}(-i\tau) \\
{\rm exp}(i\tau) &0 \\
\end{array}
\right) \label{eq:hwp1}
\end{eqnarray}

\noindent In the basis of x and y linearly polarized states,
the SU(2) matrix for $H(\tau/2)$ is given by,

\begin{eqnarray}
H(\tau/2)=-i
\left( 
  \begin{array}{lr}
{\rm cos}\tau &{\rm sin}\tau \\
{\rm sin}\tau &-{\rm cos}\tau\\
\end{array}
\right) \label{eq:hwp2}
\end{eqnarray}

\noindent From either of eqns.(\ref{eq:hwp1}) or (\ref{eq:hwp2}), it can  easily be verified that 
\begin{eqnarray}
H(\tau/2)H(\tau/2)=-
\left( 
  \begin{array}{lr}
1 & 0 \\
0 &1 \\
\end{array}
\right)
={\rm exp}(\pm i\pi)\left( 
  \begin{array}{lr}
1 & 0 \\
0 &1 \\
\end{array}
\right) \label{eq:fwp}
\end{eqnarray}

This proves the above assertion. Experimentally, this can be verified in an interference 
experiment in which a pair of identical halfwave plates is placed in one arm of a 
Mach-Zhender interferometer in 
the configuration $H(\tau/2)H(\tau/2-90^\circ)$ and the second halfwave plate is rotated  
through $90^\circ$ so that the final configuration of the pair is $H(\tau/2)H(\tau/2)$,
while no change is made in the reference arm of the interferometer. The initial Jones matrix of the 
pair is {\bf 1} and the final matrix is {\bf -1}. Consequently, for any incident polarized wave,  
the total phase shift measured will be $+\pi$ or $-\pi$ if the phase shift is continuously monitored.
Another way to see that  the phase shift is $\pm \pi$ is to realize that a pair of halfwave plates 
is equivalent to an optical rotator and apply the considerations in section 3.
How does one distinguish the topological phase shift of magnitude $\pi$ from a possible isotropic 
phase shift due to a change in optical path during rotation of the halfwave plate ? This is simple. 
If the halfwave plate were rotated through $360^\circ$ while the phase shift is being continuously 
integrated, the net phase shift due to a change in optical path must be equal to zero, whereas 
the phase shift of topological origin would integrate to $\pm 4\pi$. A related experiment that 
demonstrates this "unbounded" nature of topological phase shifts was reported in ref.\cite{unbounded}.
Another distinction between the topological and the isotropic phase in the above experiment is that 
in case of the former some polarization states get a phase $+\pi$ and some get a phase $-\pi$, while 
in case of the latter all states get the same phase.
Note that the factor $i$ multiplying the matrices in eqns.(\ref{eq:hwp1}) and (\ref{eq:hwp2}) is 
nontrivial and cannot be omitted. \footnote {When the Jones matrix of a halfwave plate is written 
without the factor $i$ as done in ref.\cite{senthil3} and in some texbooks on optics, it 
implies an isotropic phase factor ${\rm exp}(\pm i\pi/2)$ multiplying the SU(2) part. This 
must be removed before applying the considerations of this section.}

\section{Discussion}

The main issue dealt with in this paper is the phase acquired by a light wave when its 
polarization state changes under the action of a polarization transforming device. While 
the focus in this paper is on SU(2) transformations that conserve total intensity, the discussion 
applies to situations where the Jones matrices $J_1$ and $J_2$ have factors representing isotropic 
absorption and/or dichroism. In a paper written in 1956,  Pancharatnam made two 
important contributions to this problem \cite{panch1}.

(1) It was suggested that the phase difference $\delta$ between two waves in different states of polarization be 
defined such that they are in phase when the intensity resulting from their superposition is maximum. 
This leads to the definition  $\delta$=arg$({\tilde{\psi_2}}^\dagger\tilde{\psi_1})$ if we adopt the 
convenient convention that isotropic phase shifts which are not due to polarization changes are excluded 
from the definition of $\delta$. 

(2) The second result can be re-stated as follows: if the polarization state of a light wave is taken 
along a closed geodesic polygon on the Poincar\'{e} sphere by means of a sequence of transformations 
each of which takes the state along a geodesic arc on the sphere, the wave acquires a phase shift equal 
to half the solid angle subtended by the polygon at the centre of the sphere. This phase is now known 
in literature as the "geometric phase" or  "Berry's phase".

When a wave undergoes an arbitrary sequence of transformations which does not result in a  
geodesic evolution on the sphere, it acquires a total phase which is a sum of two terms: (a) the geometric 
phase equal to half the solid angle of the closed curve on the Poincar\'{e} sphere as stated in (2) above and (b) 
a dynamical phase which is determined by the evolving state and the Jones matrix of the transformation.
We have called the total phase  the "Pancharatnam phase" in this paper.  It is important to 
note that in a general evolution the geometric part of the phase  is not zero. It is a 
piece of the total phase.

In a simple evolution of the kind considered in this paper, namely a state undergoing rotation through 
a full circle about 
a fixed axis on the Poincar\'{e} sphere, the geometric phase is equal to $\pm \pi(1-{\rm cos}\theta)$ and the 
dynamical phase is equal to $\pm \pi {\rm cos}\theta$, their sum being equal to $\pm \pi$. 
While the 
decomposition of the phase in a general evolution in a geometric and a dynamical part, first done by 
Aharonov and Anandan \cite{aa}, is theoretically very interesting, the phase that is measured in an experiment 
is always the total phase. This is the reason for our choice of the total phase as the interesting 
quantity in this paper.

It also needs to be pointed out that the definition of  Pancharatnam phase requires two waves; 
one or both of which may undergo polarization changes. In a situation where the polarization state of 
only one wave changes, 
the definition of the acquired phase $\delta$ still needs a reference state and the measured phase 
depends on this reference state. In the example considered in section 2, if $J_2={\bf 1}$, i.e. only 
the wave 1 sees the transformation $R(\gamma)$, the acquired phase $\delta$ has the same behaviour 
as that shown in fig.2 with $2\gamma$ replaced by $\gamma$. Such a phase shift can be seen in a 
Mach Zhender interferometer.\\

{\large {\bf Note Added:}}

After submitting this manuscript we  became aware of the work of Tavrov et al. \cite{tavrov} in 
which they use the geometric spin redirection phase due to out of plane propagation of light  
to realise an achromatic 
$\pi$- phase shift between the two beams of an astronomical 
interferometer for "nulling interferometry". In our judgement, the linear phase shift between 
the beams for circular polarization shown in fig. 2a and the highly nonlinear phase shift 
for linear polarization shown in fig. 2b of their paper correspond approximately to the 
curves A and C shown in fig. 2 of this paper.\\


\begin{thebibliography}{99}

\bibitem{berry1}                 
M.V. Berry, ``Quantal phase factors accompanying adiabatic changes'', 
Proc. Roy. Soc. London {\bf A 392}, 45-57 (1984) .

\bibitem{panch1}
S.Pancharatnam, ``Generalized theory of interference and its applications'',
Proc. Indian. Acad. Sci. {\bf A 44}, 247-262 (1956).


\bibitem{unbounded}
R. Bhandari, ``Observation of nonintegrable geometric phase on the Poincar\'{e} sphere'',  
Phys. Lett. {\bf A 133}, 1-3 (1991) .

\bibitem{jumps}
R. Bhandari, "SU(2) phase jumps and geometric phases", Phys. Lett. {\bf A 157}, 221-225 (1991).

\bibitem{4pism}
R. Bhandari, ``4$\pi$ spinor symmetry - some new observations'', Phys. Lett. {\bf A 180}, 15-20 (1993) .

\bibitem{iwbs}
R. Bhandari, ``Interferometry without beam splitters - a sensitive technique 
for spinor phases'', Phys. Lett. {\bf A 180}, 21-24 (1993) .

\bibitem{rbdirac}
R. Bhandari, ``Observation of Dirac singularities with light polarization - I, II'', 
Phys. Lett.{\bf A 171}, 262-270 (1992) .

\bibitem{rbphysica}
R. Bhandari, ``Evolution of light beams in polarization and direction'',  Physica {\bf 175}, 111-122(1991) .

\bibitem{rbreview}
R. Bhandari, ``Polarization of light and topological phases'', Phys. Rep. {\bf 281}, 1-64 (1997).

\bibitem{chiaowu}
R.Y. Chiao and W.S. Wu, ``Manifestations of Berry's topological phase for the photon'', 
Phys. Rev. Lett. {\bf 57}, 933-936 (1986) .


\bibitem{frinsdultz}
E.M. Frins and W. Dultz, ``Direct observation of Berry's topological phase by using 
an optical fiber ring interferometer'', Opt. Comm. {\bf 136}, 354-356 (1997).

\bibitem{tomchiao}
A. Tomita and R.Y. Chiao, ``Observation of Berry's topological phase by use of an optical fiber'', 
Phys. Rev. Lett. {\bf 57}, 937-940 (1986) .


\bibitem{anandan}
J. Anandan, ``Non-adiabatic non-abelian geometric phase'', Phys. Lett. {\bf A133},  171-175 (1988).

\bibitem{wilczek}
F. Wilczek and A. Zee, ``Appearance of gauge strucures in simple dynamical systems'', 
Phys. Rev. Lett. {\bf 52}, 2111-2114 (1984).



\bibitem{senthil3}
P. Senthilkumaran, ``Berry's phase fiber loop mirror characteristics'', J. Opt. Soc. Am.-{\bf B 22}, 
505-511 (2005).

\bibitem{senthil1}
P. Senthilkumaran, G. Thursby and B. Culshaw, ``Fiber-optic tunable loop mirror using Berry's 
geometric phase'', Opt. Lett. {\bf 25}, 533-535 (2000).

\bibitem{senthil2} 
P. Senthilkumaran, G. Thursby and B. Culshaw, ``Fiber optic Sagnac interferometer for the 
observation of Berry's topological phase'', J. Opt. Soc. Am.-{\bf B 17}, 1914-1919 (2000).

\bibitem{hariharan}
P. Hariharan, "Optical Interferometry", 2nd Edition, (Academic Press 2003), p.57.


\bibitem{aa}
Y. Aharonov and J. Anandan, ``Phase change during a cyclic quantum evolution'', Phys. Rev. Lett. {\bf 58}, 
1593-1596 (1987).

\bibitem{tavrov}
A. Tavrov, R. Bohr, M. Totzeck and H. Tiziani, "Achromatic nulling interferometer based on a geometric 
spin-redirection phase", Opt. Lett. {\bf 27}, 2070-2072 (2002).
\end{thebibliography}
\end{document}